\documentclass[a4papper]{article}

\usepackage[dvips]{color}
\usepackage{graphicx}
\usepackage{subfigure}
\usepackage{graphics}
\usepackage{rotating}
\usepackage[cp850]{inputenc}
\usepackage[spanish,english]{babel}
\usepackage{amsfonts,amsbsy,amsmath}
\usepackage{epsfig}
\usepackage{multicol}
\usepackage{colortbl}
\usepackage{pstricks,pst-grad}

\usepackage{bm}
\usepackage{amssymb}
\usepackage{xspace}
\usepackage{hyperref}
\usepackage{amsmath}
\usepackage{ulem}

\usepackage{amsmath,lscape,epsfig}
\usepackage{amsfonts}
\usepackage{amsmath}
\usepackage{amssymb}
\usepackage{bm}
\usepackage{xspace}
\usepackage{hyperref}
\usepackage{ulem}

\input epsf

\newcommand{\ptt}{\ensuremath{p_{\rm{T}}}\xspace}
\newcommand{\etg}{\ensuremath{\eta _{\rm{gap}}}\xspace}

\newcommand{\de}{\ensuremath{\rm{{\delta \eta}}}\xspace}
\newcommand{\pgev}{\ensuremath{{\rm GeV}/c}\xspace}

\newcommand{\gev}{\ensuremath{{\rm GeV}}\xspace}
\newcommand{\tev}{\ensuremath{{\rm TeV}}\xspace}

\newcommand{\bc}{\ensuremath{b_{\rm{Corr}}}\xspace}
\newcommand{\FB}{\ensuremath{\rm{F-B}}\xspace}

\newcommand{\pp}{\ensuremath{\rm p\!+\!p}\xspace}

\newcommand{\auau}{\ensuremath{\rm Au\!+\!Au}\xspace}
\newcommand{\mpi}{\ensuremath{\rm{MPI}}\xspace}

\newcommand{\alice}{\ensuremath{\rm{ALICE}}\xspace}
\newcommand{\atlas}{\ensuremath{\rm{ATLAS}}\xspace}
\newcommand{\estar}{\ensuremath{\rm{STAR}}\xspace}
\newcommand{\lhc}{\ensuremath{\rm{LHC}}\xspace}

\begin{document}

\title{\bf{Extraction of multiple parton interactions and color reconnection  from  forward-backward multiplicity  correlations}}


\author{Eleazar Cuautle\thanks{ecuautle@nucleares.unam.mx},       
  Edgar Dominguez  and Ivonne Maldonado \\
  Instituto de Ciencias Nucleares, Universidad Nacional Aut\'onoma de M\'exico,\\ 
Ciudad Universitaria, Aparatado Postal 70-543, CdMx 04510,  M\'exico} \maketitle{}


\begin{abstract}
  Forward-backward multiplicity correlations have been  studied in electron-positron, proton-proton and more recently in lead-lead collisions. For the proton-proton case, comparison of experimental results to different models reveals an incomplete understanding of the physical phenomenon associated with these correlations. In this work, we present a study of forward-backward multiplicity correlations in proton-proton collisions using the PYTHIA event generator, at LHC energies. A detailed analysis is presented with and without weak decays, splitting data samples into soft and hard  QCD processes, and comparing the computed correlations for  short and long range pseudorapidity regions. Each of these regions is analyzed accounting for the effects of color reconnection and independent multiple parton interactions. We show that a combination  of these effects is required to explain the latest measurements on proton-proton data. Furthermore, is shown that from measurements of multiplicity correlations is possible to extract the average number of multiple parton interactions in the event producing these correlations, and albeit model depending, to predict the strength of these correlations, not yet measured, for higher energy collisions.
\end{abstract}

\section{Introduction}\label{intro}
The forward backward (\FB) multiplicity correlations have been studied since long time ago, for different colliding systems. One of the first results in \pp collisions in the  Intersecting Storage Ring ($\rm{ISR}$) at CERN at $\sqrt{s}=52.6$ \gev ~\cite{pplow}  was the finding of positive values for correlations. Five years later there were results of  ${\rm p+\bar{p}}$ collisions at $\sqrt{s}=540$ \gev~\cite{pantip} in the Super Proton Synchrotron ($\rm{SPS}$), also positive values of the correlation a possible dependence on the energy was reported as well. Short after, results on $e^+e^-$ collisions at $\sqrt{s}=29$ \gev~\cite{e+e-} were published and the reported results showed no \FB multiplicity correlations. This result  was interpreted as a consequence of the system studied, whereby the correlation is stronger in \pp and ${\rm p+\bar{p}}$ than in $e^+e^-$.
The E735 collaboration at the Tevatron confirmed the  dependence of the correlation with energy~\cite{ppmed} in \pp collision at $\sqrt{s}\approx 1$ \tev.
Results on heavy ions at $\sqrt{s_{NN}}=200$ \gev published by the \estar Collaboration at the Relativistic Heavy Ion Collider ($\rm{RHIC}$) found strong correlations\cite{AAStar},  for the case of the most central $\auau$ collisions, while for \pp collisions a small correlation was found. Recently
at the \lhc similar results in \pp collisions were reported by the \atlas~\cite{ppAtlas} and \alice~\cite{ppALICE} collaborations at $\sqrt{s}=0.9,2.7$ and $7$ \tev, where they observed strong correlations. Their analyses were more detailed, investigating the \ptt dependence of azimuthal and pseudorapidity distributions of \FB multiplicity correlation and argue why the \estar Collaboration could not find these results.
\FB multiplicity correlation studies are more informative when decoupled into short and long range
components ~\cite{capela1,capela2}. Short-range correlations (SRC) are localized over a small range of $\eta$, typically
up to one unit. They are induced by various short-range effects like decays of clusters or
resonances, jet and mini-jet induced correlations. Long-range correlations (LRC) extend over a
wider range in $\eta$ and originate from fluctuations in the number and properties of particle emitting sources, e.g. clusters, cut pomerons, strings, mini-jets etc. ~\cite{capela1, capela2,Nestor1,Braun}.\\
Using the \FB multiplicity correlation approach, it is possible to examine string configurations and their
interactions along the $\eta$-range, accessible in an experiment, and also to get rid of short-range
contributions coming from resonance decays, jets, etc.\\
As expected, the experimental studies of long-range rapidity correlations can give us the information about the initial stage of high energy hadronic interactions~\cite{npa810-2008}. It has been proposed that the study of the long-range \FB multiplicity correlations between two separated rapidity windows can provide a signature of the string fusion and percolation model ~\cite{Biro1984,Bialas1986,Braun1992} in ultrarelativistic heavy ion collisions~\cite{prl73-1994}.\\
The forward backward multiplicity correlations can be studied as a function of pseudorapidity ($\eta$) for windows of width \de, symmetrically around pseudorapidity $\eta =0$, and also as a function of the pseudorapidity gap (\etg) between the two \de.
Experimental results of the \FB multiplicity correlations show a common trend: they increase with increasing bin width of pseudorapidity. One obvious interpretation would be that for small bin widths the statistic fluctuations play a larger role and so they dilute the correlation strength. This asseveration has been verified by computing the \FB multiplicity correlation with a toy model.  However,  the shape of this correlation does not fit the data and there is not way to modify it. Nevertheless it is important to remember that particle production in the central pseudorapidity region is dominated by hard QCD processes and that in the forward region, soft processes increase and become more important than the hard ones. Therefore \FB multiplicity correlations are more complex than just fluctuations of particles in bin size of pseudorapidity, which are used to report the correlation. In fact, in the present work, the correlations   will be  examined for  central and forward pseudorapidity bins.  \\
Results of the \FB multiplicity correlations as a function of \etg,
both theoretical and experimentally show a decrease. Furthermore, the cut on the minimum transverse momentum of the particles in an event has been used to split events into soft and hard, as reported by \atlas~\cite{ppAtlas}, with similar results  for the \etg variable. Alternatively, in the present work, at simulation level we 
generate an analyze soft and hard QCD processes with desired conditions independently.
In order to compare our calculation in a systematic way, most of our results are presented as a function of \de with width size as reported by \alice experiment~\cite{ppALICE}. 
\\
This work is organized as follows: in section 2 we present a definition of \FB multiplicity correlations and the concepts of color reconnection (CR) and multiple parton interactions (\mpi). In section 3, the  \FB multiplicity correlations from PYTHIA are  calculated,  as well as the effects on it from weak decays, CR, the number of \mpi and combinations of them which are all accounted for to explain the experimental data. The strength of CR and the number of \mpi are extracted from data and they are used to predict \FB multiplicity correlations for energies not yet reached by the experiment. Finally, conclusions are drawn in section 4.

\section{ \FB multiplicity correlations, color reconnection and multiple parton interactions}\label{Concepts}
Multiplicity fluctuation show a  \FB multiplicity correlations defined by 
\begin{equation}
\qquad \qquad b_{Corr} (\de) =\dfrac{\langle n_{F}n_{B} \rangle-\langle n_{F} \rangle \langle n_{B} \rangle}{\langle n_{F}^2 \rangle-\langle n_{F} \rangle^2},
\end{equation}

\noindent
where $n_F$ and $n_B$ is the charged particle multiplicity in two symmetrically located \de pseudorapidity bins, separated by a central pseudorapidity gap, $\Delta \eta$.\\
The strength of the correlation factor, \bc, is sensitive to changes of multiplicity. For example, the computed values of \bc for ranges of multiplicity produce  negative values. This result is a consequence of the definition, but has no physical meaning. In the case of events with multiplicity larger than zero but limited from above, the \bc is well defined, this will be  discussed in the next section.\\
The main variable to describe the \bc is the multiplicity distribution. It is well known that the evolution of this distribution is adequately described by the negative binomial distribution~\cite{alice-mult}. In fact, in order to describe the multiplicity distribution, the number of multiple particle interactions in an event was introduced. Evidence of the importance of this variable has been reported by the experiments~\cite{exp-mpi} through  the  broadness of the multiplicity distribution. From the theoretical side, there are calculations\cite{Sjostrand87} that  have been included in the PYTHIA event generator. This research area has gained more interest given that its development~\cite{mpi-sjostrand} helps to understand the non-perturbative QCD processes and could also be related to physics at the LHC energies.\\
  In order to estimate the number of multiple parton interactions, we need to know the perturbative QCD jet cross section for parton parton interaction~\cite{Sjostrand87}:
  \begin{equation}
\frac{d\sigma}{d\ptt^2} = \sum\limits_{i,j,k} \int \int \int f_{i}^{a}(x_{1},Q^2) f_{j}^{b}(x_{2},Q^2) \frac{\hat \sigma_{i,j}^{k}}{d\hat t} \delta(\ptt^2-\frac{\hat t \hat u}{\hat s}) dx_{1}dx_{2} d\hat t
  \end{equation}
\noindent
  where $f_{i (j)}^{a (b)} (x, Q^2)$ are the parton distribution functions of the incoming partons   $i (j)$,  carrying a fraction $x$ of the energy and longitudinal momentum of the incoming hadron $a (b)$, for a given factorization and renormalization scale $Q^2 =\ptt^2  = \hat t \hat u /\hat s$,   with the hard scattering cross section ($\hat \sigma_{i,j}^{k}$)  for k-th sub-process between incoming partons $i$ and $j$, and a fragmentation function ($\delta$). The Mandelstam variables are related for massless partons by $\hat s + \hat t +\hat u = 0$.
  The hardness of the parton-parton interaction is provided by the corresponding integrated cross section which depends on the  $p_{\rm{T,min}}$ scale:
  \begin{equation}
  \qquad \qquad \sigma_{int}(p_{\rm{T,min}}) =   \int_{p_{\rm{T, min}}^2}^{s/4} \frac{d\sigma}{dp_{\rm{T}}^2} dp_{\rm{T}}^2 .
\end{equation}

  \noindent

Diffractive events  contribute with  a small fraction of the perturbative jet activity,
however, these events do not contribute to elastic processes. Therefore, the simplest 
model sets out to describe only inelastic non-diffractive events,
with an approximately known cross section. It is thus concluded that the average of such events should contain hard interactions. An average above unit corresponds to more than one sub-collision per event, which is allowed by the multiple structure of the incoming hadrons, described by the following expression~\cite{mpi-sjostrand}:
\begin{equation}
\qquad \qquad \langle n_{MPI}(p_{T, min})  \rangle = \frac{\sigma_{int}(p_{T_{min}})}{\sigma_{nd}}
\end{equation}

\noindent
where $\sigma_{nd}$ and $\sigma_{int}(p_{T_{min}})$  correspond to the cross section for non diffractive events and to the integrated one, respectively.

Of course $\langle n_{MPI} \rangle$ is multiplicity-dependent and seems to saturate according to the previous calculation~\cite{npa956-749}, increasing for forward compared to central pseudorapidities.\\
It becomes important to explore the average number of \mpi and  its relationship with  \FB multiplicity correlations since these encode 
essential information on the borderline between perturbative and non-perturbative physics, as has been discussed~\cite{mpi-sjostrand}.\\

Color reconnection (CR) could be connected to the number of  \mpi, though this represents an independent  research avenue~\cite{CR-Review}. 
Starting from  the lowest $p_T$ interaction in a set of multiple parton interactions, a reconnection probability for an interaction with hardness scale $p_T$ is given by  $P_{rec}(p_{T})$
\begin{equation}
  \qquad \qquad P_{rec}(p_{T}) =\frac{(R_{rec}\; p_{T_0})^{2}}{(R_{rec}\; p_{T_0})^{2} + p_{T}^{2}}
\end{equation}
\noindent
where the range of CR,  $0 \leq R_{rec} \leq 10$, is a phenomenological parameter and  
$p_{T_0}$  is an energy dependent parameter used to damp the low $p_{T}$ divergence of the $2 \rightarrow 2$ QCD cross section.
CR was essential to describe successfully the average transverse momentum of charged hadrons at LHC energies~\cite{pt-ALICE}. Specific applications of CR include top quark, $Z^0$ and $W^{\pm}$ decays, since they happen after previous hard perturbative activity like initial and final state radiation, as well as multiple parton interactions,  but still inside the  hadronizing color fields, thereby allowing CR with the rest of the event. For LHC studies, several new CR models were implemented in PYTHIA 8~\cite{CR-models}. Other studies on CR  have been proposed as an alternative mechanism to produce flow like effects in proton-proton collisions \cite{our-prl} where the direct variable that changes with the CR  is  the  multiplicity distribution, that decreases when CR increases and, consequently, it should produce an effect on the \FB multiplicity correlations multiplicity.

 \section{Forward-Backward multiplicity correlations with PYTHIA}\label{bcorrelation}
 The strength of the \FB multiplicity correlations (\bc)   is analyzed   with the Monte Carlo event generator PYTHIA 8.2 \cite{PYTHIA8-2}  tune Monash 2013 \cite{pskands2014}, taking into account soft and hard QCD process, with a sample of 25 billion events for each set of studies, selecting charged primary hadrons with transverse momenta in the range $0.3< p_{\rm {T}}<1.5$ \pgev and pseudorapidity range  $|\eta|<1$. These values were used  to compare our results with those reported by the  ALICE experiment~\cite{ppALICE}, and a different set of cuts were taken to  investigate other energies.
 PYTHIA Monash  2013 was used to study F-B multiplicity correlations since it is one of the last tunes which incorporates  some of the latest results of the LHC and has been used to predict \pp results at $\sqrt s =100$ \tev, indicating that up to 27 \tev, the results are quite similar with other MC event generators \cite{JHEP08-170-2016, epjC78-963-2018}.

The pseudorapidity distribution is related to the number of charged particles in an event. Therefore,  any change on this distribution, will bring consequences to the correlations. It is known that  soft  QCD processes produce lower and wider pseudorapidity distributions than the  harder ones. Then,  it is important to quantify the effects on \bc.

 \subsection{\bc and weak decays }\label{resonances}
\begin{figure}[h!]
 	\centering
 	\includegraphics[width=0.45\textwidth]{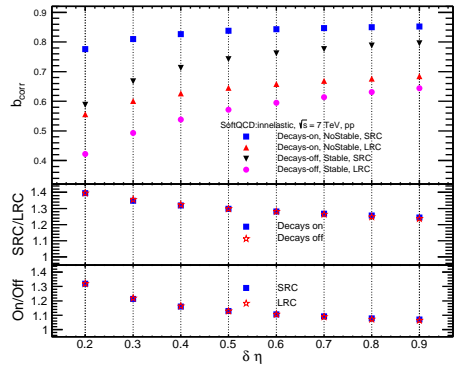}
 	\caption{Correlations with and without weak decays for short and long range correlations. The middle panel shows the ratio of \bc with to without weak decays.  Bottom panel shows the ratio of \bc for central to forward pseudorapidity multiplicity distributions.}
 	\label{fig.2}       
 \end{figure}

 Resonances, as mentioned before, allow to decouple the short and long range correlations. It is then important to look for the effects on \bc. The upper panel of Fig.~\ref{fig.2} shows the \bc for central (two upper distributions),   $|\eta|<1$ and forward pseudorapidity (two lower distributions) ranges, $3<|\eta|<4$, with and without weak decays, the last one  computed only with primary particles, as is usually reported by experiments. The \bc computed shows large effects of weak decays at low \de, where  short range correlations have larger contributions compared to long range pseudorapidity correlations where  there are not contributions from weak decays. These effects are reduced when  \de increases, as show by the ratios on the middle and the bottom panel.

\begin{figure}[h!]
   \centering
   \includegraphics[width=0.45\textwidth]{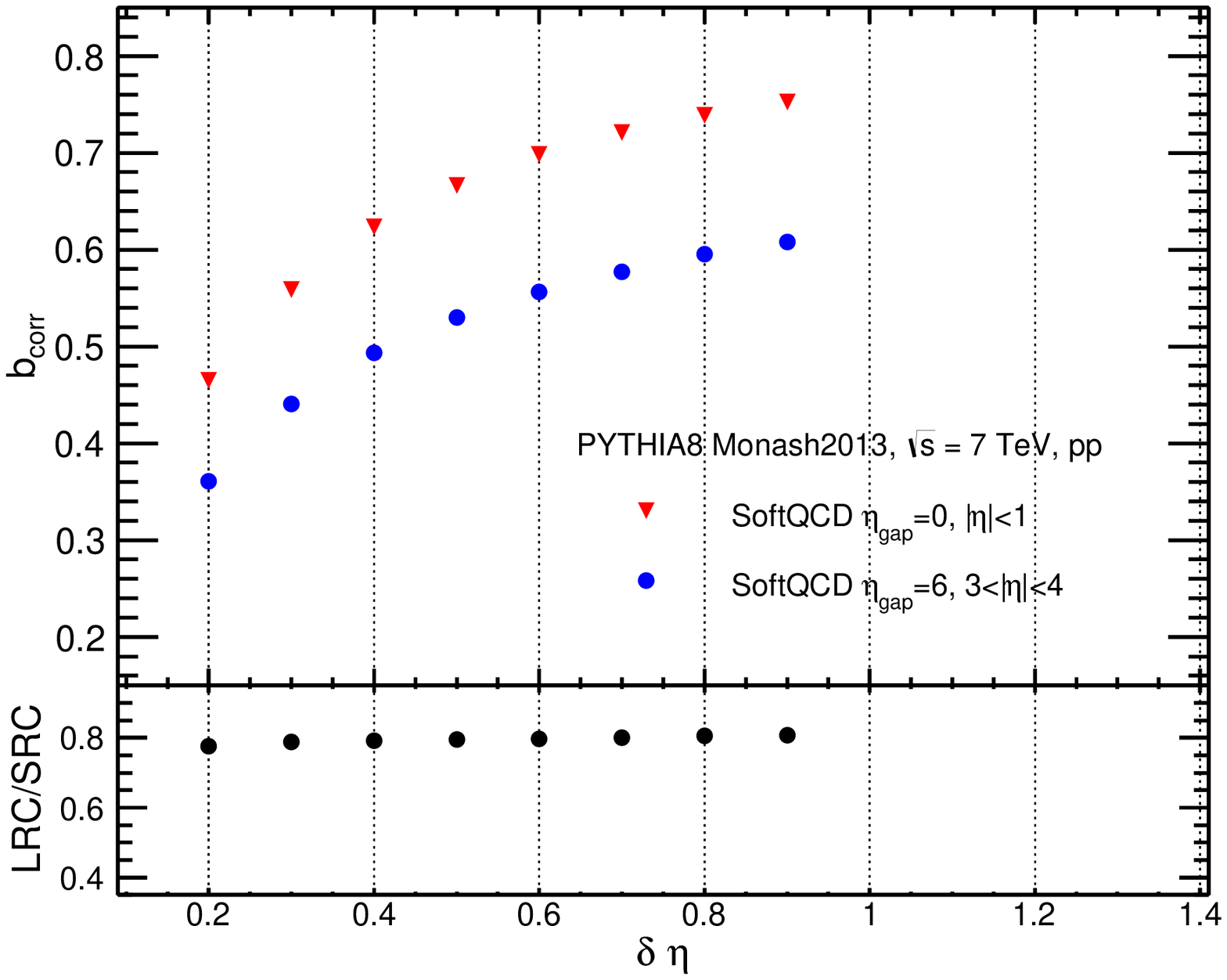}
   \includegraphics[width=0.45\textwidth]{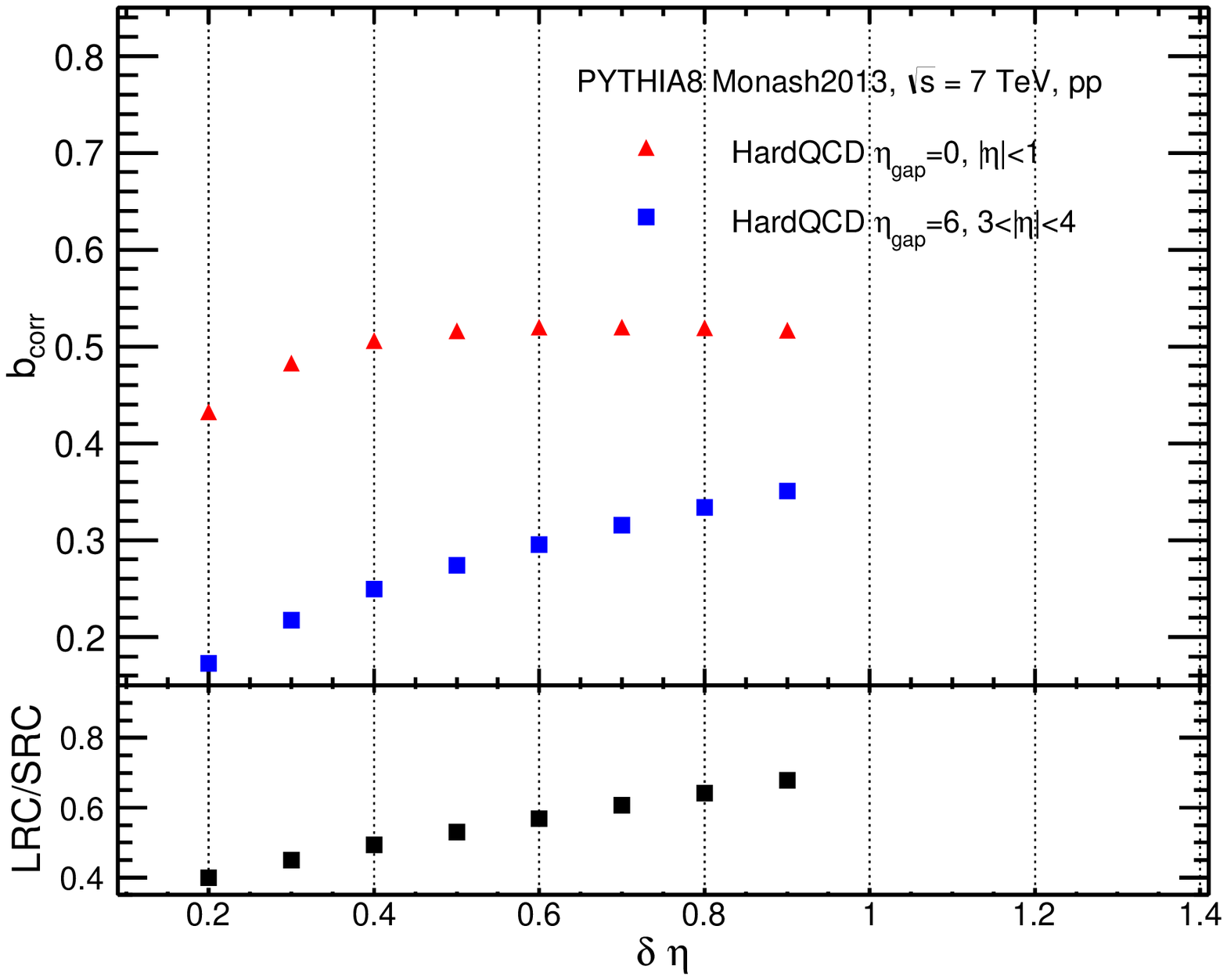}
   \caption{Correlation factor for \pp collision at $\sqrt{s}=7$ \tev for soft (upper) and hard (bottom) QCD processes,  each of them showing results for central and fragmentation pseudorapidity regions. Each of the bottom panels show the ratio LRC to SRC strength correlations}
   \label{fig.4}       
 \end{figure}

   The differences between these results raise  the possibility to explore phenomenological aspects on non-perturbative QCD processes and their effects on the \bc when the calculation is done for low \ptt. On the contrary, \bc for high \ptt particles is related to the perturbative QCD regimen and then it is possible to explore effects on  \bc from minimum bias experimental results, and or jets, for instance.\\
   Figure~\ref{fig.4} shows the \bc for soft (upper panel) and hard (bottom panel) processes, each one for two $\eta$ ranges, as indicated in the figure. Higher values are observed from soft with respect to those from hard processes. One can observe that \bc decreases  when $\eta$ goes from the central region, where short range correlation (SRC) are expected, to  large $\eta$ values were long range correlations (LRC) are expected. In each of the bottom panels the ratio  LRC to SRC is plotted, where it is observed an  almost scaling  behavior of the \bc for soft processes, while for hard processes the ratio increases. It is then important to attribute the change of the slope of \bc to hard processes. 
     Let us emphasize that for our calculations hard QCD processes are those whose transfered transverse momentum  ($\hat p_{\rm T}$)  between interacting partons is larger than 30 \gev, so that we guaranty really hard events.

 \subsection{\bc and multiple parton interactions}
\begin{figure}[h!]
   \centering
   \includegraphics[width=0.45\textwidth]{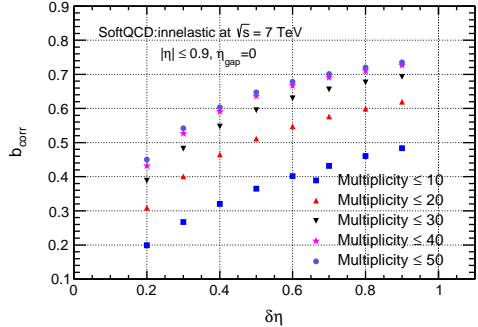}
   \caption{Correlation factor for \pp collision at $\sqrt{s}=7$ \tev, for different event classes, according to their multiplicity range.}%
   \label{fig.5}
 \end{figure}

 Since  by definition \bc  is a multiplicity dependent quantity, then it is important to identify and quantify these dependency to explore  effects on \bc. Figure~\ref{fig.5} shows \bc for different event classes according to its $N_{ch}$ range, larger than zero and less than 10, 20, 30, 40, and  50 as indicated in the figure.
  The \bc increases as the multiplicity does, this behavior does not allow to explain why the hard processes which contain more  particles in central pseudorapidity with respect to the soft ones, have lower strength in the forward backward  correlations. Nevertheless, hard events have a higher \ptt distribution which reduces the \bc, in agreement with results observed~\cite{ppAtlas} in the experiment.

\begin{figure}[h!]
   \centering
   \includegraphics[width=0.45\textwidth]{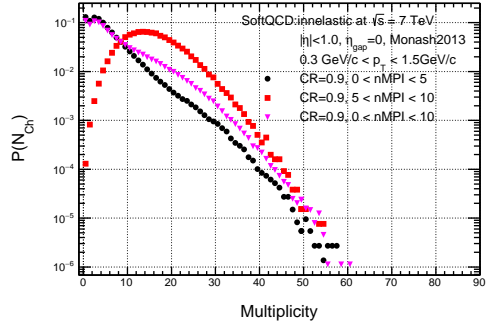}
   \caption{ Multiplicity distributions for \pp event classes  according to different number of \mpi ranges.} %
   \label{fig.6}
 \end{figure}

 In section~\ref{Concepts} we discussed the  multiplicity and how it changes with the number of multiple parton interactions. Figure ~\ref{fig.6} shows the multiplicity dependence for event classes according to  their ranges of the average number of  \mpi: $0 < \mpi < 5$, $0 < \mpi < 10$ and  $5 < \mpi < 10$, for an energy of 7 \tev. For these event classes the average multiplicities are $\langle N_{ch} \rangle$= 5,05,  6.85 and  15.24, respectively.  This illustrates that the average multiplicity distribution increases  as the  average number of  \mpi do.
   It is important to note that the average multiplicity  may increase (case of $ 5 < $  \mpi $< 10$), although the integrated  multiplicity may not.  
   Consequently, one should expect an increase of the \FB multiplicity correlation when going from low to high average number of \mpi event classes,  but not for event classes with a higher average multiplicity. This is a consequence of the selected range for the number of  \mpi.   It is worth mentioning that the strength of this correlation should saturate as a consequence of the relationship between multiplicity and the number of \mpi~\cite{npa956-749}.

\begin{figure}[h!]
   \centering
   \includegraphics[width=0.45\textwidth]{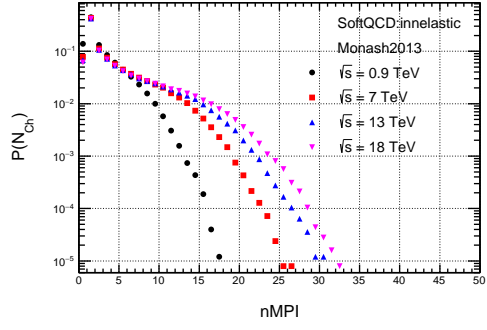}
   \caption{ Number of \mpi distributions for  \pp collisions at  $\sqrt{s}=0.9, 7, 13 $ and 18 \tev.}
   \label{fig.6.1}
 \end{figure}

   It is also essential to take into account that the distribution of number of  \mpi  as a function of the energy increases  as illustrated in the  Fig.~\ref{fig.6.1}, where a comparison 
for \pp collisions at  0.9, 7, 13 and 18 \tev is done.  Nevertheless, these distributions show a saturating trend as the energy increases.
 
\begin{figure}[ht!]
  	\centering
       \includegraphics[width=.5\textwidth]{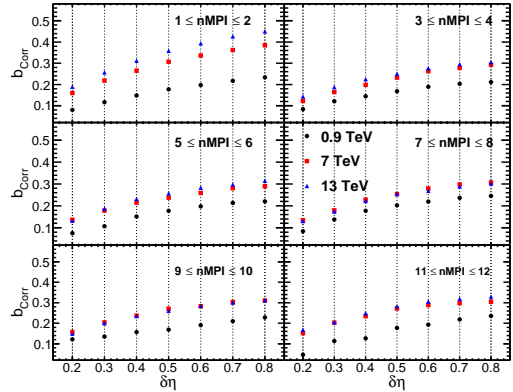}
  	\caption{  Correlation factor as function of \de for three energies and different  \mpi ranges.} 
  	\label{fig.7}      
  \end{figure}

Figure \ref{fig.7} shows the evolution of the \bc for three energies and the distinct number of \mpi ranges. Without quantifying, one can recognize small difference between 7 and 13 \tev while a large discrepancy is noticed with respect to 0.9 \tev, when these are compared in the case of the \mpi range from 1 to 2 (see the first panel of the figure). This difference may come from differences in the  number of multiple parton interactions. The discrepancy increases as \de does, however for  ranges with higher average number of \mpi   the 7 and 13 \tev cases    seem to be the same, and a scaling of the 0.9 \tev case is observed, as shown in the last panel of the figure. The examined behavior could be understood in a qualitative way from figure~\ref{fig.6.1}, which shows almost the  same trend going from 2 to 7 on the number of \mpi for all distributions.  In general, one sees a reduction of the \bc for higher ranges of the average number of \mpi as is clearly shown in the case of 7 and 13 \tev where a complete overlap is found (see the bottom panels of the second column of the figure).  This fact is a consequence of the average number of \mpi distributions, which have a similar behavior  up to $\simeq$ 12.
In addition, notice that 12 is around the saturation limit  of the number of \mpi when it is plotted versus multiplicity for 7 \tev~\cite{npa956-749}. The decreasing values of \bc in Fig.~\ref{fig.7}, going from low to high average number of \mpi intervals, could be due  to the  reduction in the multiplicity observed for the \mpi ranges used in that plot.

 Events with a low   average number of \mpi could be understood as those with a low average multiplicity~\cite{npa956-749}, compared to those  of high  average number of \mpi which are associated to  events of higher  average  multiplicity.  It is also possible to  classify those events in terms of other variables like  spherocity ($S_0$) where \bc has been studied~\cite{skand-epjc71-2011} for which the jetty-like events are those with low values of $S_0$ and low multiplicity,  and the  anisotropic ones are those with high values of $S_{0}$ and high multiplicity. In this context \bc as a function of the number \mpi could be used as an alternative variable to investigate the underlying events.

\subsection{\bc and  color reconnection}
\begin{figure}[h!]
   \centering
     \includegraphics[width=0.45\textwidth]{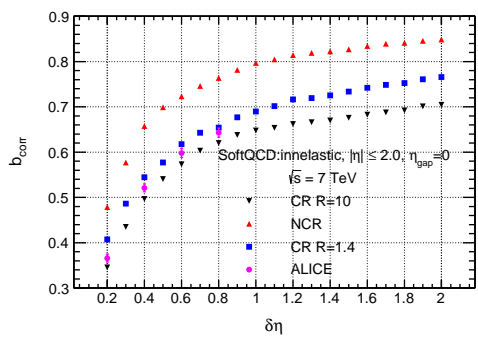}
   \caption{ \bc for  \pp collisions at $\sqrt{s}= 7$ \tev. for two strength values of CR and without it.}
   \label{fig.8}
 \end{figure}

Another crucial contribution to the \FB multiplicity  correlations is produced by the CR discussed in Sec.~\ref{Concepts}. It is well known that CR reduce the average multiplicity distribution and as a consequence, it should produce a  decrease of the \bc.
Figure ~\ref{fig.8} shows the ALICE data  compared to  \bc with and without CR where  one can observe that an increase of the strength of CR produces a decrease on the  \FB multiplicity correlations. The values used in Fig.~\ref{fig.8}  for the strength of CR are 0 (NCR in the figure),  1.4 and 10 (CR R in the figure), but none of them agrees with the data.  The correct values could be found  by an experimental data fit.

\begin{figure}[h!]
	\centering
        \includegraphics[width=.45\textwidth]{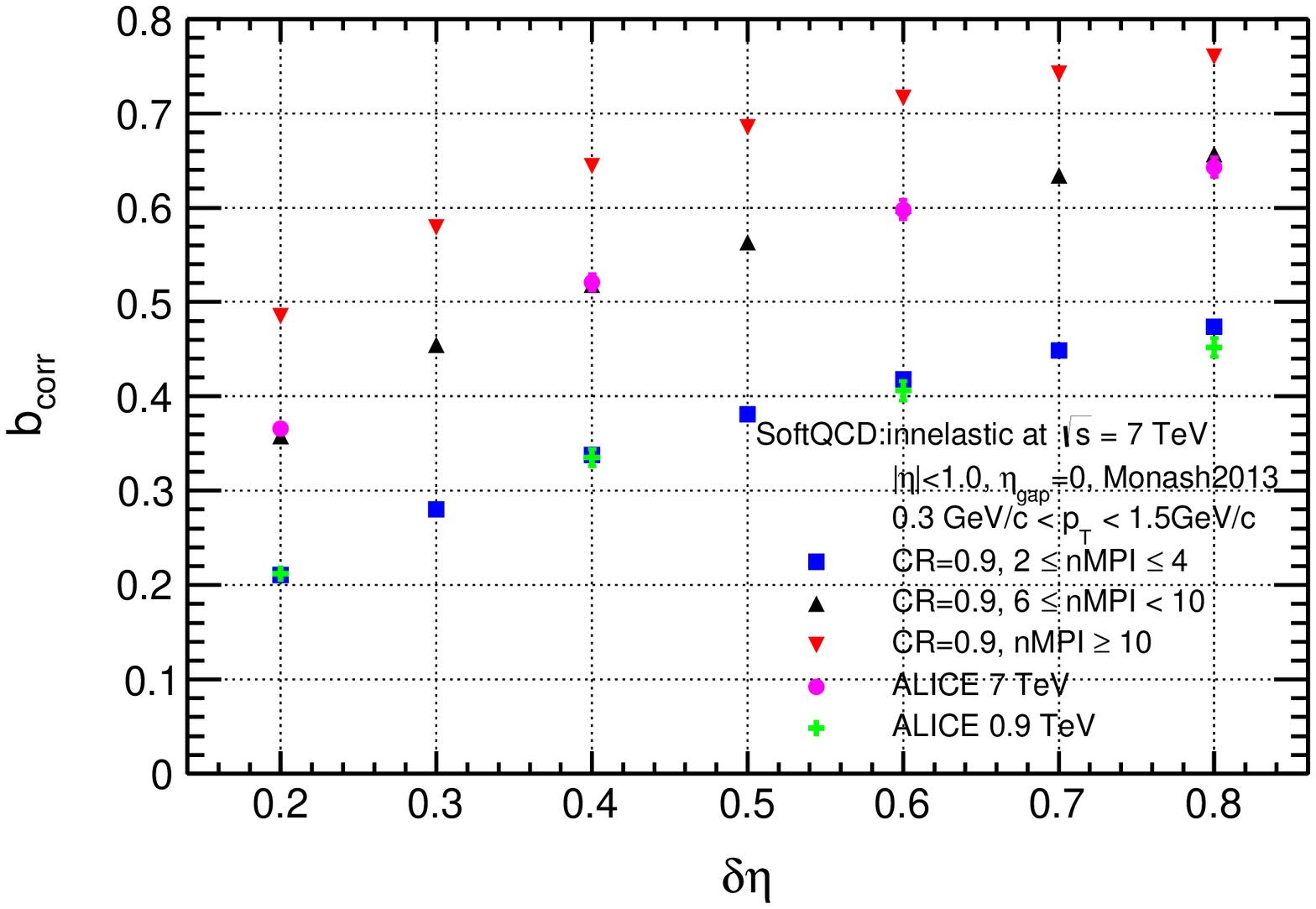}
        \includegraphics[width=.45\textwidth]{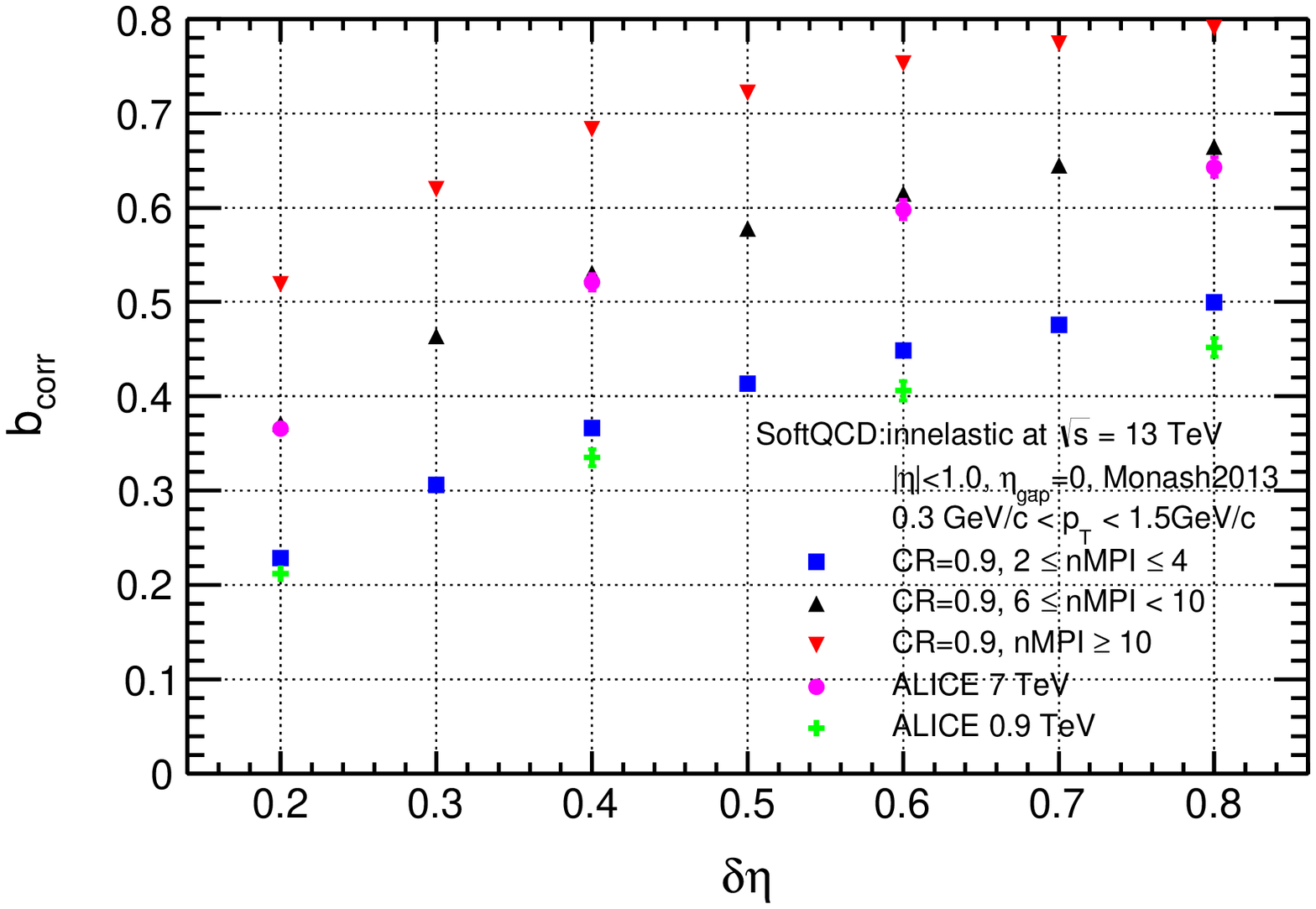}
	\caption{The correlation factor in terms of ranges of  \mpi and CR for \pp collision at $\sqrt{s}=7$ \tev (upper) and 13 \tev (bottom), compared to ALICE data at 0.9 and 7 \tev.} 
	\label{fig.9}      
\end{figure}

\noindent
This section has focused  on  the \bc and its  evolution when  CR and the number of \mpi are taken into account. Separately, an increase on the number of  \mpi produces an enhancement of the \bc, however, an increase of the strength of the CR reduces the \bc, so we need to extract simultaneously the strength of both effects. The upper panel of Fig. ~\ref{fig.9} shows our simulation of \bc at  7 \tev while the bottom one shows \bc at 13 \tev, each of them for three sets of cuts on the average  number of \mpi and CR values of 0.9, compared to data at 0.9 and 7 \tev. The best range of values  of the multiple parton interactions are  $2 \leq \mpi \leq 4$ to describe data at 0.9 \tev and $6 \leq \mpi \leq 10$ for data at  7 \tev.

Following the behavior of the multiple parton interaction distributions on  Fig.~\ref{fig.6.1}, for instance, dividing those distributions by one of them, it is possible to see the range of \mpi where they split. Furthermore, with the evolution of \bc as function of the number of \mpi of Fig.~\ref{fig.7}, one can get an insight on \bc as follows:
the upper panel of Fig.~\ref{fig.9} shows our results of the \bc at 7 \tev in the range $6\leq \mpi \leq 10$ extracted from a fit of ALICE data at 7 \tev.  It is worth noting that  the same range of the  number of \mpi but computed at 13 \tev can also reproduce data at 7 \tev (bottom panel of Fig.~\ref{fig.9}),  which is not exactly but is close to the case of data at 0.9 \tev, showed in both panels.  This number of \mpi corresponds roughly to the separation point of the distributions of Fig. ~\ref{fig.6.1} at 0.9 and 7 \tev, 7 and 13 \tev. This could be a coincidence but following  this trend of the number of \mpi distributions, \bc for the  average number of \mpi in the range:
$10 \leq \mpi \leq 12$ could correspond to data at 13 \tev. However, the plot shows results for the  average number of $ \mpi \geq 10$ which is $\sim$ 7\% higher.
This last distribution computed at 13 \tev is $\sim 3 \%$ larger than those obtained at 7 \tev. Then, considering the possibility to extract the strength of  color reconnection, for instance in average transverse momentum~\cite{pt-ALICE}, one can use the \FB multiplicity correlations strength to extract the  average  number of \mpi  event classes.
The previous procedure allows to get the average number of \mpi, and  not the number of event-by-event interactions. Furthermore, the procedure to get our results is model depend since we are using one specific multiple parton interaction model, and even more, the color reconnection is not completely independent of the multiple parton interaction models.

It is also important to  emphasize that we can make an extrapolation of the \bc to different energies and their correspondent number of \mpi. Specifically,  the ALICE data of the Fig.~\ref{fig.9} at  0.9 \tev are  fitted with a function  \bc$= a + b \; ln \sqrt (\de) $ where $a=0.49$ , $b =0.17$ and with
this parametric function scaled by a factor 1.36, one gets a correlation distribution very  close to those for data at 7 \tev. It is worth noting that the distribution of \bc for lower \mpi computed at 13 \tev presents the larger differences at higher values of \de, which can be associated to hard processes, according to the previous discussion.

 The previous results based on PYTHIA are computed incorporating color reconnection combined with multiple parton interactions to describe the data, thereby one can see the importance of the final state effects, and in general,  the phenomenology of non-pQCD reached in the experiment.  The energy dependence has also been observed, then knowledge of the relationship among the \bc, \mpi and CR brings the  possibility  to get insight into the rich phenomenology of the soft QCD process.

 \section{Conclusions}
In this work, we compared the charged hadron production in symmetric forward and backward pseudorapidity windows, in central and fragmentation pseudorapidity regions, using PYTHIA8.2 \pp event generator. The strength of this correlation, usually represented by \bc, has been studied taking into account different effects on the hadron production like weak decays,  color reconnection, multiple parton interaction, collision energy  and splitting the events into soft and hard QCD processes. Comparison with available data was also done, with the following results:\\
The general trend is that the correlations observed  at lower \de values are diluted versions of those observed for  the maximal bin size, and so provide little further discriminating power between the event samples compared here. However, slightly discrepancy noticed  between lowest  and  largest eta values could give insight  into hadron production mechanism for soft and hard QCD processes.\\
Weak decays have an important role for central rapidity; they produce an enhancement around 40\% on \bc for \de = 0.2 and  approximately 5\%  for  \de = 0.9. In the case of the fragmentation region, the behavior is similar but with higher \bc.
In general, when the resonances are  introduced, the \bc increases. This happens when the multiplicity increases, and is  higher for central than forward pseudorapidity. However, it is important to point out that the differences between \bc for central and forward pseudorapidity region decrease as \de increase, meaning that
these  correlations could be used to understand the soft and hard processes.
In fact, analyzing the soft  QCD processes  one finds a scaling between long and short range correlations. Meanwhile for hard processes, a faster saturation of the \bc  at central rapidity with respect to the fragmentation region is observed.\\
From the experimental  point of view,  the correlation strength for configurations of azimuthal sectors enables the distinction of two contributions, short  and long-range  correlations. However,  using event generators it is possible to split soft and hard QCD processes and analyze them in a separately way, as was done in the present work, to get similar conclusions.\\
Furthermore, color reconnection produces an almost constant  reduction of the strength of the \bc; in the present work the  maximum is around 14\% for all \de values. In a similar way, the collision energy produces  an almost constant  enhancement on \bc. The discrepancies could be attributed to a not linear relationship between the number of \mpi and  the multiplicity.
The number of multiple parton interactions increases with the collision energy, as well as for more isotropic events where the multiplicity increases and consequently the \bc strength increases. However it saturates as the number of \mpi does. Since the higher (lower) number of \mpi can be associated to isotropic (jetty-like) events and the \bc is  lower for jets than for isotropic events, this asseveration  is attractive and  may be used to  investigate the number of \mpi through the correlations on multiplicity in the underlying  events from the experimental point of view.\\
Finally, without trying to get a tune on the event generator, the simulation of \pp collisions at one energy and with an appropriate event selection on its number of \mpi, it is possible to reproduce the experimental data at different energies, and consequently approach the predictions to other energies, taking care that this results are dependent on the \mpi model.

\section*{Acknowledgements}
  Partial support was received by DGAPA-PAPIIT N109817 and CONACYT A1-S-16215 projects. The authors would like to  thanks A. Ayala for careful reading of the manuscript and for the useful comments.

%

\end{document}